 \documentclass[aps,prb,twocolumn,amsmath,amssymb]{revtex4-1}\newcommand{\physrevsubmit}{test} 
 
\usepackage{graphicx,color} 

\def\trans#1{#1^t}
\def\i{i}

\def\half{{1\over2}}
\def\tot{{\rm tot}}
\def\study{paper}
\def\AKLT{0}
\def\HQF{{\hat{H}}}

\def\vec#1{\mbox{\boldmath $#1$}}
\def\symop{\hat{\cal S}}
\def\ketVac{\ket{F}}
\def\ketVBS{\ket{\Phi}}
\def\ketVBSwS#1{\ket{\Phi^{(S={#1})}}}

\newcommand{\ket}[2][]{{\left|#2\right\rangle_{\!#1}}}
\newcommand{\bra}[2][]{{_{#1}\!\!\left\langle #2\right|}}
\newcommand{\braket}[3][]{{_{#1}\!\!\langle #2|#3 \rangle_{#1}}}
\def\refeq#1{Eq.~(\ref{#1})}
\def\fig#1{Fig.~\ref{#1}}
\def\tab#1{Table~\ref{#1}}
\def\sec#1{\S~\ref{#1}}
\def\Fig#1{Figure~\ref{#1}}
\def\refsec#1{\S \ref{#1}}
\def\citep#1{\cite{#1}}
\def\vv#1#2{\left(\begin{array}{c}#1\\#2\\\end{array}\right)}
\begin{document}
\title{Generalization of the Affleck-Kennedy-Lieb-Tasaki Model\\ for Quantum Ferromagnetism}

\ifdefined\physrevsubmit
\author{Isao {Maruyama}}
\email[]{i-maruyama@fit.ac.jp}
\affiliation{Department of Information and Systems Engineering, Fukuoka Institute of Technology, 3-30-1 Wajiro-higashi, Higashi-ku, Fukuoka 811-0295, Japan}
\author{Shin {Miyahara}}
\affiliation{Department of Applied Physics, Fukuoka University, 8-19-1 Nanakuma, Jonan-ku, Fukuoka 814-0180, Japan}
\date{\today}
\begin{abstract}
\else
\author{Isao {Maruyama}$^1$\thanks{i-maruyama@fit.ac.jp}, Shin {Miyahara}$^2$}
\inst{$^1$Department of Information and Systems Engineering, Fukuoka Institute of Technology, 3-30-1 Wajiro-higashi, Higashi-ku, Fukuoka 811-0295, Japan
  \\
$^2$Department of Applied Physics, Fukuoka University, 8-19-1 Nanakuma, Jonan-ku, Fukuoka 814-0180, Japan}
 
\abst{
  \fi
  We study a spin-\(S\) ferromagnetic model with exactly-written ground states, known as the partially-magnetized valence bond solid (VBS) states with magnetization $m=(S-1)/S$, which is a ferromagnetic generalization of the Affleck-Kennedy-Lieb-Tasaki model.
  We find that the VBS state and an antiferromagnetic ground state with magnetization $m=0$ are degenerate for $S=3/2$ and $S=2$ by using the Lanczos method and the density matrix renormalization group method (DMRG).
  However,  increasing $S$, the magnetization of the ground states is uniquely determined as the fraction  $m=(S-1)/S$.
  This is not just a ferromagnet, but a quantum ferromagnet due to quantum entanglement inherent in VBS states.
  In the low-energy excitation spectrum, we find the coexistence of the Haldane gap and Goldstone-like ferromagnetic magnon excitation.
  This ``magnetic chimera'' clearly appears under a finite magnetic field.
  Finally, we discuss an application to the measurement-based quantum computation and an extension of the Haldane's conjecture.

\ifdefined\physrevsubmit
\end{abstract}
\else
}
\fi
\maketitle

\section{Introduction}
Ferromagnets have been used in many industrial applications including traditional computers, and are usually expressed by fully-polarized Ising states, i.e., classical states.
Meanwhile, antiferromagnets can exhibit quantum entanglement among spins due to quantum spin fluctuation and have potential for application, for example, in quantum computing.
Both properties can coexist in ferrimagnets with antiferromagnetic coupling\cite{SR.10.9193, JPSJ.90.064707, JPSJ.63.2359, SSC.98.245, PRB.91.155136, NPJQM.9.96, NC.13.6888, JPSJ.67.3711, JPCM.10.11033, PRB.59.1024, PRB.65.214403}; spontaneous symmetry breaking in ferrimagnetism due to mixed spins is explained by the Lieb-Mattis theorem\cite{JMP.3.749}.
Recently proposed ``quantum ferromagnet''\cite{PRB.108.140404, PRB.110.014433} can also exhibit the coexistence despite single spins without translational symmetry breaking; spontaneous magnetization satisfies the Oshikawa-Yamanaka-Affleck criterion\cite{PRL.78.1984}.

\begin{figure}
  \begin{center}
  \includegraphics[width=0.35\textwidth]{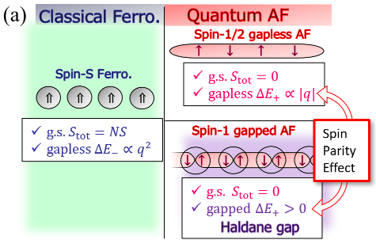}\\
  \includegraphics[width=0.35\textwidth]{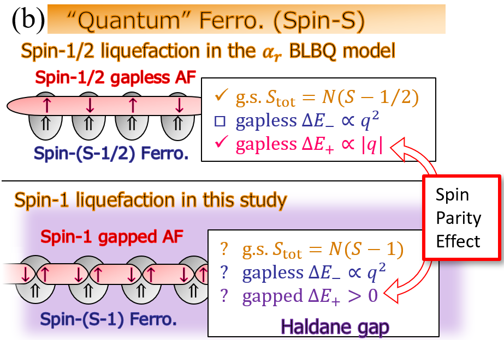}
  \end{center}
  \caption{(Color online)Schematic picture of the ground states (a) in traditional spin-$S$ chain models and (b) in ``quantum'' ferromagnetism of spin-$S$ chain models.
    Properties about total spin $S_\tot$ of the ground states and low energy excitation $\Delta E_\pm$ are also summarized.
    For $\Delta E_+$, the spin-1/2 antiferromagnet (AF) has $\Delta E_+\propto |q|$, i.e., des Cloizeaux-Pearson mode\cite{PR.128.2131} while the spin-1 AF has the Haldane gap $\Delta E_+>0$\cite{PRL.50.1153}.
    On the other hand, $\Delta E_-\propto q^2$ is Goldstone-type gapless one-magnon mode\cite{ISBN.0521551439}.
  }
\label{figV}
\end{figure}
The schematic pictures of the conventional ferromagnets and antiferromagnets in one-dimensional Heisenberg models are illustrated in \fig{figV}~(a). 
In this \study, the spins in classical ferromagnets are depicted separately, whereas  the spins in quantum antiferromagnets are connected with neighboring spins to illustrate the existence of quantum entanglement.
The quantum ferromagnet in the spin-$S$ bilinear biquadaratic (BLBQ) model induced by spin-1/2 liquefaction is shown in the top panel of \fig{figV}~(b), where one can combine a spin-$(S-1/2)$ classical ferromagnet and a spin-1/2 quantum antiferromagnet by using rigorous ``eigensystem embedding'', i.e., the exact eigensystem correspondence between the spin-$S$ BLBQ model and the spin-1/2 Heisenberg model\cite{PRB.108.140404}.
The rigorous correspondence, which might be interesting in the context of quantum many-body scars\cite{NP.14.745, PRL.119.030601, PRB.98.235155, PRB.98.235156, PRL.124.180604}, is limited to the specific point $\alpha_r$ in the spin-$S$ BLBQ model but is valid on any lattice in any dimension.
Then, interesting research topics for quantum spin-1/2 antiferromagnets, for example, solvable models\cite{ZP.71.205}, spin-liquid states\cite{NATURE.464.199, SCIENCE.332.1173}, and resonating-valence-bonds (RVBs) \cite{SCIENCE.235.1196}, may be embedded in quantum ferromagnets for a large enough spin $S$.

In fact, the ferromagnetic Haldane phase was discovered\cite{PRB.110.014433} by applying the theory to the well-known antiferromagnetic Haldane phase in the spin-1/2 ladder\cite{EPJB.15.227}.
The numerical calculation on the dynamical structure factor $S^\pm(q,\omega)$, one of experimental observables, theoretically predicts the co-existence of gapless mode $\Delta E_-$ and gapped mode $\Delta E_+$.
As a ferromagnetic property, $\Delta E_-\propto q^2$ is expected in spite of small system size\cite{note1}.
As an antiferromagnetic property, $\Delta E_+$ is quantitatively identical to that of the spin-1 Heisenberg model\cite{PRB.48.3844}, i.e., the Haldane gap, $\Delta E_+>0$.

The nature of the Haldane gap\cite{PRL.50.1153} in antiferromagnets has been revealed by the Affleck-Kennedy-Lieb-Tasaki (AKLT) model\cite{PRL.59.799}.
Despite nonintegrability of the AKLT model, the ground state is exactly written, and it is called the valence bond solid (VBS) state, as depicted in the right bottom panel in \fig{figV}~(a), and it has recently been referred to as the AKLT state in the context of measurement-based quantum computation (MBQC)\cite{PRL.106.070501, PRL.114.247204}.
The Haldane phase with entangled gapped quantum spin-liquid states \cite{RMP.89.040502} is a notable example of the symmetry protected topological (SPT) phases\cite{PRB.80.155131, PRB.81.064439}.
A partially magnetized VBS state, proposed by Oshikawa\cite{JPCM.4.7469}, can be a unique ground state under a magnetic field as a magnetization plateau state\cite{PRL.78.1984}.
However, because one of the degenerated ground states has a total spin $S_\tot=0$ under a zero magnetic field, their model is not appropriate for ferromagnets.
Unique total-spin $S_\tot$ of all the ground states is needed to ensure spontaneous magnetization under a zero magnetic field, which is an important aspect of ferromagnets.

In this \study, to reveal the nature of the ferromagnetic Haldane phase, we extend the spin-\(\frac12\) liquefaction\cite{PRB.108.140404} to spin-1 liquefaction in the following order.
In \sec{sec:VBS}, we define a ferromagnetic AKLT state [the bottom panel of \fig{figV}~(b)] following Oshikawa\cite{JPCM.4.7469} and we obtain a matrix-product state (MPS) form.
We define a ferromagnetic AKLT Hamiltonian, whose ground state is supposed to be the ferromagnetic AKLT state in \sec{sec:H}, and we give an analytical proof of the ground states in \sec{sec:proof}.
In \sec{sec:results}, we present the numerical results of the Lanczos method and the density matrix renormalization group method (DMRG)\cite{PRL.69.2863, RMP.77.259} to answer the three questions in the bottom panel in \fig{figV}~(b).
As a potential application, we consider the effects of a finite magnetic field to obtain a unique and gapped ground state in \sec{sec:Eb} and apply the ground state to MBQC in \sec{sec:MBQC}.
We present the summary in \sec{sec:summary}.

\section{Ferromagnetic AKLT States\label{sec:VBS}}
\begin{figure}
  \begin{center}
  \includegraphics[width=0.3\textwidth]{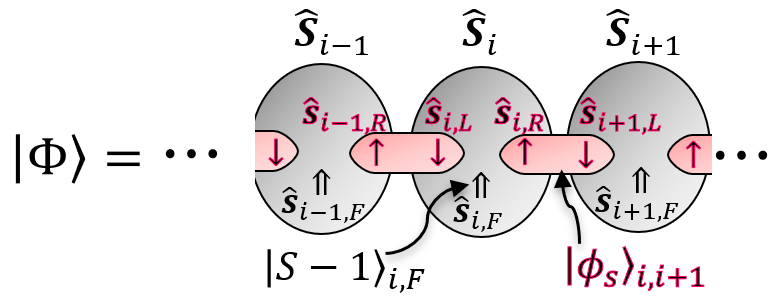}\\
  \end{center}
  \caption{(Color online)Ferromagnetic AKLT state $\ketVBS$ in \refeq{VBS} written with spin-singlets $ \ket{\phi_s}_{i,i+1}$ and background ferromagnetic Ising states $\ket{S-1}_{i,F}$.  Spin-$S$ operator $\vec{\hat{S}}_i$ at the $i$-th site is decomposed into spin-$(S-1)$ operator $\vec{\hat{S}}_{i,F}$ and two spin-1/2 operators $\vec{\hat{s}}_{i,L}$, $\vec{\hat{s}}_{i,R}$.
    This is identical to Oshikawa's state\cite{JPCM.4.7469} and is depicted in the bottom panel of \fig{figV}~(b).
  }
\label{figVM}
\end{figure}
We consider one-dimensional spin-$S$ models.
To define a ferromagnetic AKLT state, let us divide spin-$S$ operator $\vec{\hat{S}}_i$ at the $i$-th site into three kinds of decomposed spins, $\vec{\hat{s}}_{i,L}+\vec{\hat{s}}_{i,R}+\vec{\hat{s}}_{i,F}$: left spin-1/2 operator $\vec{\hat{s}}_{i,L}$, right spin-1/2 operator $\vec{\hat{s}}_{i,R}$, and front spin-$(S-1)$ operator $\vec{\hat{s}}_{i,F}$, as depicted in \fig{figVM}.
The spin-$S$ ferromagnetic AKLT state $\ketVBS$ in one dimension under the periodic boundary condition (PBC) is defined, using symmetrization mapping operator $\symop$, as
\begin{eqnarray}
  \ketVBS &=& \symop \prod_{i=1}^N \ket{\phi_s}_{i,i+1} \ket{S-1}_{i,F}
  \label{VBS}
\end{eqnarray}
with the spin-singlet state 
\begin{eqnarray}
  \ket{\phi_s}_{i,j}= \ket{+1/2}_{i,R}\ket{-1/2}_{j,L} - \ket{-1/2}_{i,R}\ket{+1/2}_{j,L}
\end{eqnarray}
on the bond $i,j$.

The state $\ketVBS$ is identical to Oshikawa's ferromagnetic Ising-VBS state\cite{JPCM.4.7469},
where ``ferromagnetic Ising'' means a ferromagnetically ordered Ising-state $\prod_{i} \ket{S-1}_{i,F}$ for the spin-$(S-1)$ sub-system $\vec{\hat{s}}_{i,F}$.
The term ``VBS'' comes from an analogy to the valence bond theory for a covalent bond in chemistry\cite{BUSSEIKENKYU.58.121}; each bond-singlet state $\ket{\phi_s}_{i,i+1}$ is formed by two spin-(1/2)s, which participate from the $i$-th and $(i+1)$-th spin respectively, i.e., $\vec{\hat{s}}_{i,R}$ and $\vec{\hat{s}}_{i+1,L}$.
For the $S=1$ case,
\begin{eqnarray}
 \ketVBSwS{1}= \symop \prod_{i=1}^N \ket{\phi_s}_{i,i+1},
  \label{VBSwS1}
\end{eqnarray}
obtained by omitting $\ket{S-1}_{i,F}$ in \refeq{VBS}, is identical to the spin-$1$ VBS ground state\cite{PRL.59.799,BUSSEIKENKYU.58.121}.
In other words, $\ketVBS$ is a natural generalization of the spin-$1$ case.

Because the local spin-singlet $\ket{\phi_s}_{i,j}$ can be written in a quadratic form with the two-dimensional matrix:
\begin{math}
  \ket{\phi_s}_{i,j}
=
    \left(\ket{+\half}_{i,R}\; \ket{-\half}_{i,R}\right)
\left(
    \begin{array}{cc}
      0 & 1
      \\
      -1 & 0
    \end{array}
  \right)
\left(
    \begin{array}{c}
      \ket{+\half}_{j,L}\\
      \ket{-\half}_{j,L}
    \end{array}
  \right)
\end{math}
, one can deduce an MPS form of the singlets
$\prod_{i=1}^N\ket{\phi_s}_{i,i+1}= \mathrm{Tr} \prod_{i=1}^N$
$ \left(
    \begin{array}{c}
      \ket{+\half}_{i,L}\\
      \ket{-\half}_{i,L}
    \end{array}
  \right)
  $
$  \left(\ket{+\half}_{i,R}\; \ket{-\half}_{i,R}\right)$
$  \left(
    \begin{array}{cc}
      0 & 1
      \\
      -1 & 0
    \end{array}
  \right)
$ $=$
$\mathrm{Tr} \prod_{i=1}^N$
$ \left(
    \begin{array}{cc}
      -\ket{+\half}_{i,L}\ket{-\half}_{i,R} & \ket{+\half}_{i,L}\ket{+\half}_{i,R}
      \\
      -\ket{-\half}_{i,L}\ket{-\half}_{i,R} & \ket{-\half}_{i,L}\ket{+\half}_{i,R}
    \end{array}
  \right)
$ by using the trace of the two-dimensional MPS, $\mathrm{Tr}$, that comes from the PBC.
After some calculations on the symmetrization mapping operator $\symop$ in \refeq{VBS},
one can obtain the MPS form
\begin{eqnarray}
  \ketVBS&=&\mathrm{Tr}\prod_{i=1}^N A_i
\end{eqnarray}
with
\begin{eqnarray}
  A_i&=&
 \left(
  \begin{array}{cc}
    -{\ket{S-1}_i\over \sqrt{2S}}&\ket{S}_i \\
    - {\ket{S-2}_i \over \sqrt{S(2S-1)}} &{\ket{S-1}_i\over \sqrt{2S}} \\
  \end{array}
  \right)
  \label{MPS:A}
  .
\end{eqnarray}
For the $S=1$ case, 
\begin{eqnarray}
  A_i^{(S=1)}&=&
 \left(
  \begin{array}{cc}
    -{\ket{0}_i\over \sqrt{2}}&\ket{1}_i \\
    - {\ket{-1}_i } &{\ket{0}_i\over \sqrt{2}} \\
  \end{array}
  \right)
\end{eqnarray}
is equal to that of the previous study\cite{AP.326.96} except for a normalization constant; also for $S=3/2$ case\cite{PRB.91.155136}.

In addition, it is easy to obtain the matrix product operator (MPO)\cite{PRB.95.035129} form
\begin{eqnarray}
  \ketVBS &=& \left(\mathrm{Tr}\prod_{i=1}^N {\hat{A}}_i \right) \prod_{i=1}^N \ket{S}_i
              \label{MPO}
\end{eqnarray}
with
\begin{eqnarray}
   {\hat{A}}_i  &=&
         \left(
  \begin{array}{cc}
    -{\hat{S}^-_i\over 2S}&1 \\
    - {(\hat{S}_i^-)^2\over 2S(2S-1)} &{\hat{S}_i^-\over 2S} \\
  \end{array}
  \right)
\end{eqnarray}
by using $\hat{S}^-_i\ket{m}_i=\sqrt{(S+m)(S-m+1)}\ket{m-1}_i$.
This MPO, $\mathrm{Tr}\prod_i {\hat{A}}_i$, is a creation operator of the state $\ketVBS$, where a fully-saturated ferromagnetic state $\ketVac=\prod_i \ket{S}_i$ plays a role of the vacuum state: i.e., $\left(\mathrm{Tr}\prod_{i=1}^N {\hat{A}}_i \right)^\dagger \ketVac=0$.

\section{Model Hamiltonians\label{sec:H}}
\subsection{AKLT Hamiltonians}
Before we define ferromagnetic AKLT Hamiltonians, we review the one-dimensional $S=1$ AKLT Hamiltonian\cite{PRL.59.799} and related models. The one-dimensional $S=1$ AKLT Hamiltonian is defined as
\begin{eqnarray}
  \hat{H}_{\AKLT}^{(S=1)}
  \;=\;\sum_{i=1}^N {3 \vec{\hat{S}}_i\cdot \vec{\hat{S}}_{i+1}    +    \left(\vec{\hat{S}}_i\cdot \vec{\hat{S}}_{i+1}\right)^2    +    2\over 6}
  \label{def:H_AKLT:SdS}
,
\end{eqnarray}
which corresponds to the specific point $\alpha=\arctan(1/3)$ of the general BLBQ Hamiltonian $\sum_{i} [\cos\alpha \; \vec{\hat{S}}_i\cdot \vec{\hat{S}}_{i+1} + \sin\alpha (\vec{\hat{S}}_i\cdot \vec{\hat{S}}_{i+1})^2] $.
The coefficient of the bilinear term ${3\vec{\hat{S}}_i\cdot \vec{\hat{S}}_{i+1}\over 6}$ is an antiferromagnetic-type coefficient(i.e., $3/6=1/2 > 0$).
The biquadratic term $(\vec{\hat{S}}_i\cdot \vec{\hat{S}}_{i+1})^2$ with the positive coefficient $1/6$ is artificial in the sense that a negative coefficient is natural\cite{PRL.83.4176}.
This artificial Hamiltonian is important because both the ground states of $\hat{H}_{\AKLT}^{(S=1)}$ and the spin-1 Heisenberg model $\sum_i \vec{\hat{S}}_i\cdot \vec{\hat{S}}_{i+1}$ are in the same SPT phase, i.e., the Haldane phase.

The coefficients given as exact fractional values in \refeq{def:H_AKLT:SdS} come from projection operators, $\hat{P}_{ij}^{(s)}$, onto the subspace with total spin $s$ on bond $i,j$.
In the projection operator form, $\hat{H}_{\AKLT}^{(S=1)}$ is written as
\begin{eqnarray}
  \hat{H}_{\AKLT}^{(S=1)}&=&\sum_{i=1}^N \hat{P}^{(2)}_{i,i+1}
  \label{def:H_AKLT:P}
,
\end{eqnarray}
which can be proved using a general relation\cite{PRL.60.531}:
\begin{eqnarray}
  &&\hat{P}_{ij}^{(s)}= \prod_{\substack{n=0 \\ n\neq s}}^{2S} {\vec{\hat{S}}_i\cdot \vec{\hat{S}}_j- q_n \over q_s -q_n}
  ,
  \nonumber\\
  &&(\vec{\hat{S}}_i\cdot \vec{\hat{S}}_j)^n = \sum_{s=0}^{2S} {q_s}^n \hat{P}_{ij}^{(s)}
  ,
  \label{eq:AAH}
\end{eqnarray}
with $q_s ={s(s+1)\over 2} -S(S+1)$ for spin-$S$.

For one-dimensional models, a higher spin-$S$ generalization has been studied\cite{JSP.52.627}:
\begin{eqnarray}
  \hat{H}_{\AKLT}^{(S)}=\sum_{i=1}^N \hat{P}^{(2S)}_{i,i+1}
  \label{def:H:P2S}
  .
\end{eqnarray}
Even though the ferromagnetic AKLT state $\ketVBS$ is a zero-energy ground state of $\hat{H}_{\AKLT}^{(S)}$, other states with a different total spin $S_\tot$ are also zero-energy degenerated ground states for $S \geqq 3/2$.
For $S=3/2$, to stabilize the target state $\ketVBSwS{3/2}$ an infinitesimal magnetic field is required\cite{PRL.78.1984}, where the projection operator $\hat{P}^{(3)}_{i,j}$ in $\hat{H}_{\AKLT}^{(S=3/2)}$ has an additional ``bicubic'' (BC) term $(\vec{\hat{S}}_i\cdot \vec{\hat{S}}_{j})^3$:
\begin{eqnarray}
  \hat{P}^{(3)}_{i,j} = {27 \vec{\hat{S}}_i\cdot \vec{\hat{S}}_{j}\over 160}
  +{29 (\vec{\hat{S}}_i\cdot \vec{\hat{S}}_{j})^2\over 360}
  +{(\vec{\hat{S}}_i\cdot \vec{\hat{S}}_{j})^3\over 90}
  +{11\over 128}
  \label{eq:P3}
.
  \;\;\;\;\;
\end{eqnarray}
Despite the absence of unique total spin in one dimension, the spin-$3/2$ BLBQBC Hamiltonian $\hat{H}_{\AKLT}^{(S=3/2)}$ on the two-dimensional honeycomb lattice has a unique ground state with $S_\tot=0$\cite{PRL.59.799,CMP.115.477}, which has been referred to as the two-dimensional AKLT state in the context of MBQC\cite{PRL.106.070501, PRL.114.247204}.

\subsection{Ferromagnetic AKLT Hamiltonians}
To realize unique $S_\tot=N(S-1)$ ground states in one dimension under a zero magnetic field,
we define a general spin-$S$ ``ferromagnetic'' AKLT Hamiltonian:
\begin{eqnarray}
  \HQF^{(S)}=
  \sum_{i=1}^N \left[ J_{i}^{(2S)} \hat{P}^{(2S)}_{i,i+1} + \sum_{s=0}^{2S-4} J_{i}^{(s)} \hat{P}^{(s)}_{i,i+1} \right]
       ,
  \label{eq:FV}
\end{eqnarray}
with positive coefficients $J_{i}^{(s)} > 0$ for all $s$ and all $i$, which is identical to the form proposed by Tu and Sanz\cite{PRB.82.104404}.
In other words, the absence of terms with $J_{i}^{(2S-1)}, J_{i}^{(2S-2)}, $ and $J_{i}^{(2S-3)} $ is important for the definition.
Due to the completeness relation $\sum_{s=0}^{2S} \hat{P}_{i,i+1}^{(s)}=1$,
we can define a more general Hamiltonian $\hat{H}^{(S)}=\sum_{i}\sum_{s=0}^{2S}  J_{i}^{(s)} \hat{P}^{(s)}_{i,i+1}$ with the condition $J_{i}^{(2S-1)}=J_{i}^{(2S-2)}=J_{i}^{(2S-3)} < \mathrm{min}(J_{i}^{(2S)}, J_{i}^{(2S-4)}, J_{i}^{(2S-5)}, \ldots, J_{i}^{(1)}, J_{i}^{(0)})$ and with the energy shift $\sum_i J_{i}^{(2S-1)}$.

For numerical calculations, we limit ourselves to a uniform and simple Hamiltonian.
To simplify the general Hamiltonian, one can eliminate higher $n$-th order terms $(\vec{\hat{S}}_i \cdot \vec{\hat{S}}_{j})^n$ ($n \geq 5$) by properly choosing free parameters $J_{i}^{(s)}>0$.
Then, a simple Hamiltonian is given as
\begin{widetext}
 \begin{align}
   \HQF^{(S)}(\beta)=
   J \sum_{i=1}^N
   (\vec{\hat{S}}_i\cdot \vec{\hat{S}}_{i+1}+\beta)
   (\vec{\hat{S}}_i\cdot \vec{\hat{S}}_{i+1}-S^2+2S)
   (\vec{\hat{S}}_i\cdot \vec{\hat{S}}_{i+1}-S^2+4S-1)
   (\vec{\hat{S}}_i\cdot \vec{\hat{S}}_{i+1}-S^2+6S-3)
   \label{def:H_FV}
     ,
 \end{align}
\end{widetext}
which is a bilinear biquadratic bicubic biquartic (BLBQBCBQ) Hamiltonian.
Here, $J>0$ is an energy scale parameter and $\beta$ is the remaining free parameter that must satisfy the condition
\begin{eqnarray}
 -S^2 < \beta < -S^2 + 8 S-6
  \label{eq:beta_}
\end{eqnarray}
to satisfy the positivity $J_{i}^{(s)}>0$.
The lower bound of $\beta$ comes from $\epsilon_F^{(S)}(\beta) >0$, where $\epsilon_F^{(S)}(\beta)= {\langle \hat{H}^{(S)}(\beta) \rangle_F\over N}$ is the eigenenergy per site for the fully-saturated ferromagnetic state $\ketVac=\prod_i \ket{S}_i$.
The explicit form is calculated as 
\begin{eqnarray}
  \epsilon_F^{(S)}(\beta) = 6 J (S^2+\beta)S(4S-1)(2S-1) 
  \label{def:epsilon_F}
\end{eqnarray}
by substituting $\vec{\hat{S}}_i\cdot \vec{\hat{S}}_{i+1}$ with $S^2$ in \refeq{def:H_FV}.

For $S = 1$ and $S=3/2$, the Hamiltonian $\HQF^{(S)}(\beta)$ is reduced as follows,
\begin{eqnarray}
  &&
     \HQF^{(S=1)}(\beta)
  =
     \epsilon_F^{(S=1)}(\beta)\;\; \hat{H}_{\AKLT}^{(S=1)}
     ,
  \\
  &&\HQF^{(S=3/2)}(\beta)=
     \epsilon_F^{(S=3/2)}(\beta)\;\; \hat{H}_{0}^{(S=3/2)}
.
\end{eqnarray}
The coefficients comes from the ferromagnetic energies $\langle \hat{H}^{(S)} \rangle_F = N \epsilon_F^{(S)}(\beta)$ and $\langle \hat{H}_{\AKLT}^{(S)} \rangle_F =N$.
Thus, $\HQF^{(S)}(\beta)$ could be considered to be a direct generalization of $\hat{H}_{0}^{(S=1)}$ and $\hat{H}_{0}^{(S=3/2)}$. 
In addition, for $S=1/2$, we have a reduced form $\HQF^{(S=1/2)}(\beta)= 0$ which does not correspond to the general Hamiltonian $\HQF^{(S=1/2)}=\sum_i J_i^{(1)} \hat{P}^{(1)}_{i,i+1}$; however, this is because there is no $\beta$ satisfying $-1/4<\beta<-9/4$, which is \refeq{eq:beta_}.
In general, for $S\geq 2$, the biquartic term $(\vec{\hat{S}}_i\cdot \vec{\hat{S}}_{i+1})^4$ cannot be reduced in \refeq{eq:FV}.
This is equivalent to the fact that the biquadratic term $(\vec{\hat{S}}_i\cdot \vec{\hat{S}}_{i+1})^2$ in \refeq{def:H_AKLT:P} cannot be reduced for $S\geq 1$.
In this sense, the BLBQBCBQ Hamiltonian $\HQF^{(S)}(\beta)$ is essential for $S\geq 2$.

\subsection{Simplification for DMRG}
The Hamiltonian $\HQF^{(S)}(\beta)$ written in $\sum_i \prod_{k=1}^4 (a_k \vec{\hat{S}}_i \cdot \vec{\hat{S}}_{i+1}+ b_k)$ is complex but suitable for the Lanczos method.
It is enough to program an operator-times-vector routine for a general operator $\alpha \vec{\hat{S}}_i \cdot \vec{\hat{S}}_{j} + \beta$, where the output vector is $\vec{v}_{out}=(\alpha \vec{\hat{S}}_i \cdot \vec{\hat{S}}_{j} + \beta)\vec{v}_{in}$ for an input vector $\vec{v}_{in}$.
By repeating this routine four times, we obtain $\vec{v}_{out}=\prod_{k=1}^4 (a_k \vec{\hat{S}}_i \cdot \vec{\hat{S}}_{i+1}+ b_k)\vec{v}_{in}$.
After the summation of the local Hamiltonians, we obtain  $\vec{v}_{out}=\HQF^{(S)}(\beta)\vec{v}_{in}$.
However, for the DMRG code, $\HQF^{(S)}(\beta)$ is too complex.
Thus, we need a simpler Hamiltonian with a properly chosen free-parameter $\beta$.

A Hamiltonian that is suitable for the DMRG code is realized for
\begin{eqnarray}
  \beta=3(S-1)(S-3)=:\beta_S
  \label{eq:beta_S}
,
\end{eqnarray}
where the function $\beta_S=3(S-1)(S-3)$ is defined for all $S$ but the positivity condition \refeq{eq:beta_} is valid only for $2\leq S\leq 4$.
In the following, we explain simplification by $\beta_S$.
For the DMRG code, the Hamiltonian written with SU($k+1$) generators $\vec{\hat{Q}}^{(k)}_i$ is suitable.
The Hamiltonian $\HQF^{(S)}(\beta)$ in \refeq{def:H_FV} for any $\beta$ is rewritten as
\begin{eqnarray}
  \HQF^{(S)}(\beta) = \sum_{k=1}^4 C_k(\beta) \vec{\hat{Q}}^{(k)}_i\cdot \vec{\hat{Q}}^{(k)}_j
\end{eqnarray}
with coefficient $C_k(\beta)$ and the operators
\begin{eqnarray}
  {\hat{\vec{Q}}}^{(k)}_i\cdot {\hat{\vec{Q}}}^{(k)}_j
  =
  \sum_{m=0}^k 
  {{\hat{Q}}^{m+;(k-m)z}_i \left({\hat{Q}}^{m+;(k-m)z}_j\right)^\dagger
  +{\rm h.c.}
  \over 2}
,  \nonumber
  \\
  {\hat{Q}}^{m+;nz}_i
  =\sum_{l=0}^n {c_{l;m,n} \left[ (\hat{S}^+_i)^m (\hat{S}_i^z)^n + (\hat{S}_i^z)^n (\hat{S}_i^+)^m \right]\over 2} 
.\nonumber
\end{eqnarray}
Note that $\vec{\hat{Q}}^{(1)}_i=\vec{\hat{S}}_i$ for $k=1$.
One important point is that the Hamiltonian $\HQF^{(S)}(\beta_S)$ has no third-order term $\vec{\hat{Q}}^{(3)}_i \cdot \vec{\hat{Q}}^{(3)}_{j}$, i.e., $C_3(\beta_S)=0$.
As a result, the DMRG code for  $\HQF^{(S)}(\beta_S)$ requires only 10=2+3+5 operators: two SU(2)-operators $\hat{S}_i^z=\hat{Q}_i^{0+;1z}$ and $\hat{S}_i^+=\hat{Q}_i^{1+;0z},$ three SU(3)-operators $\hat{Q}_i^{0+;2z}, \hat{Q}_i^{1+;1z},$ and $\hat{Q}_i^{2+;0z}, $ and five SU(5)-operators $\hat{Q}_i^{0+;4z}, \hat{Q}_i^{1+;3z}, \hat{Q}_i^{2+;2z}, \hat{Q}_i^{3+;1z},$ and $\hat{Q}_i^{4+;0z}$.
The Hamiltonian's coefficient $C_k(\beta)$ and coefficients $c_{l;m,n}$ of ${\hat{Q}}^{m+;nz}_i$ will be detailed in elsewhere.

\subsection{Summary of Models}
Before we move on to next section, we summarize the models defined in \refsec{sec:H}.
All the Hamiltonians defined in this section are SU(2) symmetric, i.e., rotational symmetric in the spin space.
Then, the total spin $S_\tot$ and its $z$-component $S_\tot^z$ are good quantum numbers to label eigenstates.
Here, the total spin operator $\vec{\hat{S}}_\tot$ is defined as
\begin{eqnarray}
  \vec{\hat{S}}_\tot = \sum_{i=1}^N \vec{\hat{S}}_i
  .
\end{eqnarray}
Because a ferromagnetic state with the maximum total spin $S_\tot=N S$ is an excited state due to $\hat{P}^{(2S)}_{i,i+1}$,
$S_\tot < N S $ is expected for the ground states.
In addition, unlike the AKLT Hamiltonian  $\hat{H}_{0}^{(S)}$ in \refeq{def:H:P2S}, the additional terms $\hat{P}^{(0)}_{i,i+1}, \hat{P}^{(1)}_{i,i+1}, \ldots \hat{P}^{(2S-4)}_{i,i+1}$ in the general Hamiltonian $\HQF^{(S)}$ may lift up the small $S_\tot$ states among degenerated ground states in $\hat{H}_{0}^{(S)}$.
Then, the realization of ground states having a unique $S_\tot=N(S-1)$ is naively expected.

The BLBQBCBQ Hamiltonian $\HQF^{(S)}(\beta_S)$ at $\beta = \beta_s$ for $2 \leq S \leq 4$ are summarized in \tab{tab:coeff} with $\hat{H}_{0}^{(S=1)}$ and $\hat{H}_{0}^{(S=3/2)}$.
Here, the coefficients of bilinear terms are negative (ferromagnetic) for $S\geq 2$ while those for $S<2$ are positive (antiferromagnetic).
Due to the ferromagnetic bilinear terms for $S\geq 2$, ferromagnetic states are favorable but due to the positive biquartic term (${\hat{h}_i}^4$) the maximum total-spin states are not favorable.
This is a naive understanding of fractional magnetization under a zero magnetic field.
\begin{table}[ht]
  \centering
  \caption{Hamiltonian $\hat{H}_{0}^{(S=1)}$, $\hat{H}_{0}^{(S=3/2)}$, and $\HQF^{(S\geq 2)}(\beta_S)$ normalized by the coefficient of the bilinear term, $\hat{h}_i=\vec{\hat{S}}_i\cdot \vec{\hat{S}}_{i+1}$.
  }
  \label{tab:coeff}
  \begingroup
  \renewcommand{\arraystretch}{1.8}
  \begin{tabular}{|r|l|}\hline
    Hamiltonian & normalized Hamiltonian with $\hat{h}_i=\vec{\hat{S}}_i\cdot \vec{\hat{S}}_{i+1}$
    \\\hline
    $\hat{H}_{0}^{(S=1)}$& $\sum_i\left(
           {\hat{h}_i}
           +{{\hat{h}_i}^2\over 3}
           +{2\over 3}
           \right)
           $
    \\
    $\hat{H}_{0}^{(S=3/2)}$&
             $\sum_i\left(
             {\hat{h}_i}
             +\frac{116 {\hat{h}_i}^2}{243}
             +\frac{16 {\hat{h}_i}^3}{243}
             +\frac{55}{108}
             \right)
           $
    \\
    $\HQF^{(S= 2)}(\beta_2)$&
           $\sum_i\left(
           -{\hat{h}_i}
           -\frac{{\hat{h}_i}^2}{5}
           +\frac{{\hat{h}_i}^3}{9}
           +\frac{{\hat{h}_i}^4}{45}
             \right)
           $
    \\
    $\HQF^{(S= 5/2)}(\beta_{5/2})$&
             $\sum_i\left(
             -{\hat{h}_i}
             -\frac{178   {\hat{h}_i}^2}{503}
             +\frac{80 {\hat{h}_i}^3}{503}
             +\frac{16 {\hat{h}_i}^4}{503}
             +\frac{11385}{8048}
             \right)
$
    \\
    $\HQF^{(S= 3)}(\beta_{3})$&
           $\sum_i\left(
           -{\hat{h}_i}     
           -\frac{{\hat{h}_i}^2}{3}
           +\frac{5 {\hat{h}_i}^3}{36}
           +\frac{{\hat{h}_i}^4}{36}
             \right)
           $
    \\
    $\HQF^{(S= 7/2)}(\beta_{7/2})$&
             $\sum_i\left(
             -{\hat{h}_i}
             -\frac{134 {\hat{h}_i}^2}{717}
             +\frac{80 {\hat{h}_i}^3}{2151}
             +\frac{16 {\hat{h}_i}^4}{2151}
             -\frac{2415}{3824}
             \right)
             $
    \\
    $\HQF^{(S= 4)}(\beta_{4})$&
           $\sum_i\left(
           -{\hat{h}_i}
           -\frac{73 {\hat{h}_i}^2}{293}
           +\frac{5 {\hat{h}_i}^3}{293}
           +\frac{{\hat{h}_i}^4}{293}
           +\frac{360}{293}
             \right)
           $
    \\\hline
  \end{tabular}
\endgroup
\end{table}
\section{Analytical Proof\label{sec:proof}}
Our first task in this section is to prove that the ferromagnetic AKLT state $\ketVBS$ defined in \sec{sec:VBS} is a zero-energy ground state for the general Hamiltonian $\HQF^{(S)}$ of \refeq{eq:FV} in \sec{sec:H}:
\begin{eqnarray}
  \HQF^{(S)}\ketVBS=0
 \label{eq:zes}
  .
\end{eqnarray}
Because $\HQF^{(S)}$ is composed of positive semidefinite operators $\hat{P}^{(s)}_{i,i+1}$, the lowest energy is non-negative.
Then, \refeq{eq:zes} means that $\ketVBS$ is not only a zero energy eigenstate but also a ground state.
The following are two proofs for \refeq{eq:zes}, but both proofs are simple: just a two-site problem.
That is, $\HQF^{(S)}$ is frustration free.
We are considering PBC but it is easy to consider an open boundary.

The proof based on \refeq{VBS} is natural.
For two neighbor sites $i$ and $i+1$, because there exist spin ${\bf 0}$ for $\ket{\phi_s}_{i,i+1}$, spin ${\bf 2S-2}$ for $\ket{S-1}_i\ket{S-1}_{i+1}$, and a pair of free spin-1/2s for $\vec{\hat{s}}_{i,L}$ and $\vec{\hat{s}}_{i+1,R}$, the spin composition of decomposed spins becomes ${\bf 0}\otimes ({\bf 2S-2}) \otimes {\bf 1/2} \otimes {\bf 1/2}=({\bf 2S-1})\oplus({\bf 2S-2})\oplus ({\bf 2S-2})\oplus ({\bf 2S-3})$.
Then, the state  $\ketVBS$ becomes a zero-energy eigenstate of the projection Hamiltonian without $\hat{P}^{(2S-1)}_{i,i+1}, \hat{P}^{(2S-2)}_{i,i+1}$, or $\hat{P}^{(2S-3)}_{i,i+1}$.
This is a natural proof for \refeq{eq:zes}.

The other proof based on the MPS form in \refeq{MPS:A} is straight-forward.
The proof is composed of two parts, $\hat{P}_{i,i+1}^{(s)}A_i A_{i+1}=0$ $(0\leq s\leq 2S-4)$ and $\hat{P}_{i,i+1}^{(2S)}A_i A_{i+1}=0$, where  matrix product $A_i A_{j}$ becomes
\begin{math}
  A_i A_{j} =
\end{math}
\begin{math}
  \left(
    \begin{array}{cc}
      {(a\hat{S}^-_i- \hat{S}^-_{j})\hat{S}_{j}^-\over 2S(2S-1)} &{-\hat{S}^-_i+\hat{S}^-_{j}\over 2S} \\
      {\hat{S}_i^- (\hat{S}^-_i-\hat{S}^-_{j})\hat{S}_j^-\over 4S^2(2S-1)} &{\hat{S}_i^- (- \hat{S}_i^-+a \hat{S}^-_j)\over 2S(2S-1)}\\
    \end{array}
  \right)
\end{math}
\begin{math}
\ket{S}_i\ket{S}_{j}
\end{math}
with $a={2S-1\over 2S}$.
The four states in the MPS $A_iA_{j}$ have $S_{i,j}^z=S^z_i+S^z_{j}\geq 2S-3$, which gives the lower bound of total spin as $S_{i,j} \geq 2S-3$ for the four states; thus, we obtain $\hat{P}_{i,j}^{(s)}A_i A_{j}=0$  $(0\leq s\leq 2S-4)$.
The remaining task is to prove $\hat{P}_{i,j}^{(2S)}A_i A_{j}=0$.
First, let us classify the four states in $A_iA_{j}$ with parity for the swap of indices $i,j$;
three elements, $(A_i A_{j})_{1,2}$, $(A_i A_{j})_{2,1}$, and $(A_i A_{j})_{1,1}-(A_i A_{j})_{2,2}$, have odd parity, whereas the remaining state $(A_i A_{j})_{1,1}+(A_i A_{j})_{2,2}$ has even parity.
The former three states with odd parity cannot have total spin $2S$, because the highest total spin $2S$ state must have even parity.
For the latter state with even parity, we must calculate that $(A_i A_{j})_{1,1}+(A_i A_{j})_{2,2}\propto  \ket{S-2}_i\ket{S}_{j} - \sqrt{2a}\ket{S-1}_i\ket{S-1}_{j} + \ket{S}_i\ket{S-2}_{j}
$
is orthogonal to the total spin $S_{i,j}=2S$ state with the same $S_{i,j}^z=2S-2$  written as $(\hat{S}^-_i+ \hat{S}^-_{j})^2\ket{S}_i\ket{S}_{j}\propto \ket{S-2}_i\ket{S}_{j} + \sqrt{2/a} \ket{S-1}_i\ket{S-1}_{j} + \ket{S}_i\ket{S-2}_{j}$.

The other ground states are written as $(\hat{S}_{\tot}^-)^s \ketVBS$ with $\hat{S}_{\tot}^- = \sum \hat{S}_{i}^-$.
The eigenequation $\HQF^{(S)} (\hat{S}_{\tot}^-)^s \ketVBS = 0$ is  derived  as $\HQF^{(S)} (\hat{S}_{\tot}^-)^s \ketVBS
=\hat{S}_{\tot}^- \HQF^{(S)} (\hat{S}_{\tot}^-)^{s-1} \ketVBS
=\cdots = (\hat{S}_{\tot}^-)^s \HQF^{(S)} \ketVBS=0$ 
by using
\begin{eqnarray}
 [\HQF^{(S)}, \hat{S}_{\tot}^\pm] =0 
.
\end{eqnarray}
Here, $\HQF^{(S)} \ketVBS=0$ is \refeq{eq:zes}.
Using the property\cite{note2} of the symmetrization mapping operator $\symop$
\begin{eqnarray}
  \hat{S}^-_{i}\symop = \symop\left(\hat{s}^-_{i,L}+\hat{s}^-_{i,R}+\hat{s}^-_{i,F}\right)
  \label{eq:symop}
\end{eqnarray}
for \refeq{VBS}, one can show
\begin{eqnarray}
  (\hat{S}_{\tot}^-)^s \ketVBS  =\symop \prod_{i=1}^N \ket{\phi_s}_{i,i+1} (\hat{s}_{\tot,F}^-)^s \prod_{i=1}^N\ket{S-1}_{i,F}
\end{eqnarray}
because $(\hat{s}^-_{i,R}+\hat{s}^-_{i+1,L})\ket{\phi_s}_{i,i+1}=0$.
Here, $\hat{s}_{\tot,F}^- = \sum \hat{s}_{i,F}^-$. 
It should be noted that $(\hat{s}^-_{N,R}+\hat{s}^-_{1,L})\ket{\phi_s}_{N,1}=0$ due to PBC.
Since the ferromagnetic background state $(\hat{s}_{\tot,F}^-)^s \prod_{i=1}^N\ket{S-1}_{i,F}$ has a total spin $s_{\tot,F}=N(S-1)$ and $z$-component $s^z_{\tot,F}=N(S-1)-s$,
the ground state $(\hat{S}_{\tot}^-)^s \ketVBS$ has the same total spin $S_{\tot}=N(S-1)$ and the same $z$-component $S^z_{\tot}=N(S-1)-s$.
Here, $s$ is in the range $[0,2N(S-1)]$; i.e., the number of degeneracy is $2N(S-1)+1$.

For $S=1$, the number of degeneracy is $1$ and $S_{\tot}=0$: i.e., the unique ground state $\ketVBSwS{1}$ in \refeq{VBSwS1}.
In other words, the above discussion is a simple generalization of previous studies for $S=1$.
Then, as a future research topic, one could consider anomalous features of the $S=1$ system even in the ferromagnetic AKLT model, for example, correlation functions including the string order parameters\cite{PRB.40.4709, JPCM.4.7469}, hidden $Z_2\times Z_2$ symmetry revealed by the Kennedy-Tasaki transformation\cite{PRB.45.304}, high dimensions\cite{PRL.59.799}, large spin VBS states\cite{PRL.60.531, JMP.64.031901}, recent topological indices\cite{PRL.121.140604, CMP.374.705, AX.2407.17041}, SPT for larger $S$\cite{PRB.91.155136}.
Another future work is topological classification between the ferromagnetic AKLT models and known materials, for example, ferrimagnetism in the verdazyl-based salts\cite{SR.10.9193, JPSJ.90.064707} and $m=1/3$ plateau\cite{JPSJ.63.2359, SSC.98.245, PRB.91.155136, NPJQM.9.96} in antiferromagnet Na$_2$Cu$_3$Ge$_4$O$_{12}$\cite{NC.13.6888}.

Despite the rigorous proof of the ground states, there is a possibility that another ground state than $(\hat{S}_{\tot}^-)^s \ketVBS$ exists.
Then, we need numerical results to show that the ferromagnetic AKLT state is a unique ground state in the next section.
\section{Numerical Results\label{sec:results}}
In this section, we present the numerical results of using the Lanczos method and DMRG to answer the three questions in the bottom panel of \fig{figV}~(b).
Because the simplified Hamiltonian $\HQF^{(S)}(\beta_S)$ still has both translational symmetry and SU(2) symmetry, each state is labeled by quantized numbers: wave number $q=2\pi n/N$, total spin $S_\tot$, and its $z$ component $S_\tot^z$.
Here, $n$ is an integer and $N$ is the system size.
The $S_\tot$ and $S_\tot^z$ can be calculated both in the Lanczos method and DMRG, whereas the wave number $q$ can be calculated only in the Lanczos method.
With both methods, the total spin is calculated as 
\begin{eqnarray}
  S_\tot &=& {\sqrt{4 \langle \vec{\hat{S}}_\tot\cdot \vec{\hat{S}}_\tot \rangle+1}-1\over 2}
  \label{ev:Stot}
,
\end{eqnarray}
which becomes a non-integer if states having a different $S_\tot$ are degenerated during the $S_\tot^z=0$ sector calculations.
In other words, the non-integer $S_\tot$ provides numerical evidence of the degeneracy.
Since the ground state $\ketVBS$ has $S_\tot=N(S-1)$ as shown in \sec{sec:proof},
we also define the shift in the total spin as
\begin{eqnarray}
  \Delta S=S_\tot - N(S-1)
.
\end{eqnarray}

The magnetization $m$ of a state is given by the expectation value of $S_\tot$ as
\begin{eqnarray}
  m&=& {S_\tot \over N S}
.
\end{eqnarray}
For example, the fully saturated value, $m=1$, is obtained by the fully-polarized ferromagnetic Ising state $\prod_{i=1}^N \ket{S}_{i}$.
In the spin-1/2 BLBQ chain\cite{PRB.108.140404,PRB.110.014433}, fractional magnetization $m={S-1/2\over S}={2S-1 \over 2S}$ under a zero magnetic field has been numerically observed: $m={S-1/2\over S}$ corresponds to $S_\tot=N(S-1/2)$ due to spin-1/2 liquefaction.

In addition, with using DMRG, to obtain a specific $S_\tot=n$ states for a given $n$, we consider the specific $S_\tot^z=n$ sector of a Hamiltonian $\hat{H}_\alpha = \HQF^{(S)}(\beta_S) + \alpha  \vec{\hat{S}}_\tot\cdot \vec{\hat{S}}_\tot$ for $\alpha>0$ which can lift large $S_\tot$ states.
Due to the SU(2) symmetry of the Hamiltonian, a desired $S_\tot=S_\tot^z=n$ state becomes a ground state of $\hat{H}_\alpha$ for a large enough $\alpha$ in an $S_\tot^z=n$ subspace.
Then, we calculate the energy of the original Hamiltonian for a given $n$ by using the ground state of $\hat{H}_\alpha$ as
\begin{eqnarray}
  E_{S_\tot=n}&=&
                  \langle\; \HQF^{(S)}(\beta_S)\; \rangle_\alpha 
  \label{ev:E_Stot}
.
\end{eqnarray}

We perform the DMRG finite-size method under the periodic boundary condition using a ladder configuration with and without the conserved value $S_{\tot}^z$.
The maximum number of sweeps is 10, i.e., 10 right and left sweeps in the finite-size method.
The number of remaining basis in the block is $\chi\leq 800$, where the maximum memory is approximately 100GB. 

\subsection{Unique Total Spin of Ground States}
In this subsection, we discuss whether the ferromagnetic AKLT states with $S_{\tot} = N(S-1)$ are unique ground states with the $(2S_{\tot} +1)$-fold degeneracy or not for $\HQF^{(S)}(\beta_S)$, as defined in \sec{sec:H}.

As a numerical result, the ferromagnetic AKLT states with $S_\tot=N(S-1)$ are unique under the finite-size gap for $5/2 \leq S \leq 4$ but are not unique for $S=2$.
In the latter case, the expectation value of $S_\tot$ becomes a non-integer value due to the degeneracy, for both the ED and the DMRG results in the $S_\tot^z=0$ sector.
To increase $S$ is inevitable.
Although the upper limit, $S\leq 4$, is introduced in our calculation through the specially chosen $\beta_S$ in \refeq{eq:beta_S}, it is naively expected that the uniqueness of $S_\tot$ will hold for $S>4$ with using $\beta$ in \refeq{eq:beta_}.

\begin{figure}
  \begin{center}
  \includegraphics[width=0.52\textwidth]{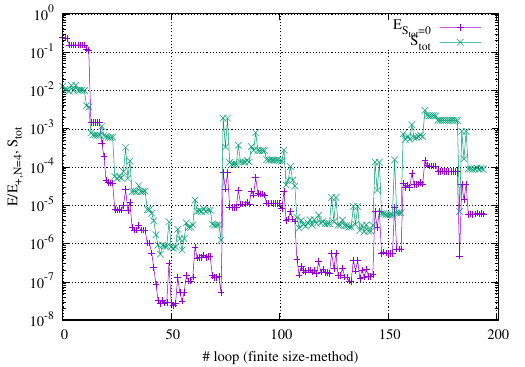}\\
  \end{center}
  \caption{(Color online)$E_{S_\tot=0}$ and $S_\tot$ as a function of the number of sweeps in the finite size method for $S=2$, and $N=24$ obtained by DMRG with $\chi=100$.
    All data points are positive: $E_{S_\tot=0} > 0$ and $S_\tot>0$.
    The unit of energy $E_{+,N=4}$ will be given by \refeq{def:scale} in the next section \sec{sec:Gp}.
  }
\label{figS2}
\end{figure}

In addition, we conclude that another ground state for $S=2$ has $S_\tot=0$.
To obtain direct evidence of the zero energy ground state with $S_\tot=0$ by DMRG, we calculate the energy $E_{S_\tot=0}$ by using the projection method in \refeq{ev:E_Stot} by sweeping $\alpha$.
The typical results for the system size $N=24$ in the finite-size-method sweeps are shown in \fig{figS2}.
Despite accumulation of numerical error after reaching the lowest energy within 3 sweeps, all data points are positive: $E_{S_\tot=0} > 0$ and $S_\tot >0$.
The positivity of energy $E_{S_\tot=0} > 0$ comes from the variational principle holding at every step of the finite-size DMRG method.
Then, \fig{figS2} shows the numerical evidence of the zero energy ground state with $S_\tot=0$ for the system size $N=24$ with a very small error $<10^{-6}$.

We briefly discuss the $S=1$ and $S=3/2$  cases.
For the $S=1$ case, the ground state  $\ketVBSwS{1}$ in \refeq{VBSwS1} with $S_\tot=0$ is unique, whereas for the $S=3/2$ case, an infinitesimal magnetic field is required to stabilize the ground state with $S_\tot=S_\tot^z=N/2$\cite{PRL.78.1984}.
We also calculate the $S=3/2$ and $N=12$ case and discover another ground state with $S_\tot=0$.

To conclude this subsection, the finite-size numerical calculation has shown that the ground states have 
\begin{eqnarray}
  S_\tot = N(S-1), \;\;\; S\neq 3/2, 2
\end{eqnarray}
uniquely.
This conclusion is an answer to the first question in \fig{figV}~(b).
Due to the total-spin rotational symmetry under a zero magnetic field, the number of degenerated ground states is $2 S_\tot+1=2 N(S-1)+1$.
The spontaneous symmetry breaking, as in the case of a ferromagnetic Heisenberg chain, can occur at zero temperature in one-dimensional models following the Mermin-Wagner theorem\cite{PRL.17.1133, PR.158.383}.
Then, the fractional magnetization $m=(S-1)/S$ is realized as a natural extension of the antiferromagnetic spin-1 AKLT chain with $m=0$ for $S=1$.
In the remaining part of this \study, we focus on $5/2 \leq S \leq 4$ and we ignore $S=3/2, 2$ due to the lack of unique $S_\tot$ of the ground states.

\subsection{Haldane Gap\label{sec:Gp}}
In this subsection, we generalize the concept of the Haldane gap to the ferromagnetic case, where the ground states have $S_\tot = N(S-1)$ uniquely.
It is expected that a triplet excitation of the spin-singlets $\prod_i \ket{\phi_s}_{i,i+1}$ in the ground state, \refeq{VBS}, can define the Haldane gap.
More precisely, the Haldane gap is defined as the lowest energy $E_+$ in the sector $S_\tot^z = N(S-1)+1$.
Because the ground states have zero energy, the lowest excitation energy $\Delta E_+$ from the ground state energy becomes
\begin{eqnarray}
  \Delta E_+ = E_+ - 0 = E_+
  \label{def:DeltaE+}
,
\end{eqnarray}
which corresponds to the $\Delta E_+$ depicted in \fig{figV}.
For $\hat{H}_{\AKLT}^{(S=1)}$,  by using the single mode approximation (SMA)\cite{PRL.60.531}, a triplet dispersion $E_{SMA}(q)={5 \over 27}(5+3\cos q)$ has been obtained; the wave number $q=\pi$ gives the lowest energy $E_{SMA}(\pi)={10 \over 27}>0$, which is identified as the Haldane gap $\Delta E_+=E_{+,q=\pi}$.

As a result for $5/2 \leq S\leq 4$, the lowest-energy state in the sector $S_\tot^z = N(S-1)+1$ has a gapped excitation energy $\Delta E_+ >0$ and a total spin $S_\tot= N(S-1)+1$, ($\Delta S=+1$).
Based on the Lanczos method, the lowest-energy state has wave number $q=\pi$ with quadratic dispersion around $q=\pi$ as in the case of $S=1$.
In addition, the number of degeneracy for the states with the lowest excitation energy $\Delta E_+$ is $2S_\tot+1=N(S-1)+3$.
For $S=1$, the number of degeneracy becomes $2S_\tot+1 = 3$, which are the spin-triplet excitations.

As an analytical result for $N=4$, obtained using Mathematica, we found the explicit formula
\begin{eqnarray}
  E_{+,N=4}={\epsilon_F^{(S)}(\beta_S) \over  (4-1/S)}
  \label{def:scale}
,
\end{eqnarray}
where $\epsilon_F^{(S)}$ is ferromagnetic energy per site in \refeq{def:epsilon_F}.
We use $E_{+,N=4}$ as a unit of energy in this \study.

\begin{figure}
  \includegraphics[width=0.52\textwidth]{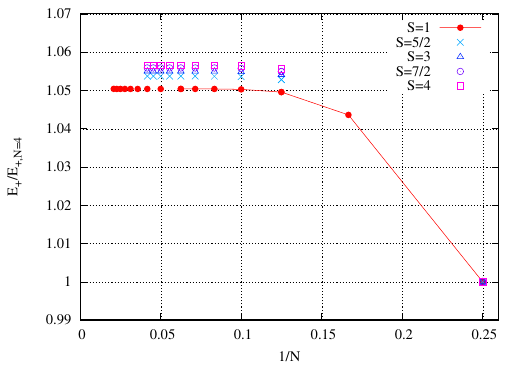}
  \caption{(Color online)System-size $N$ dependence of the Haldane gap $\Delta E_+=E_+$ of $\HQF^{(S)}(\beta_S)$ defined in \refeq{def:DeltaE+} with the unit of energy $E_{+,N=4}$ in \refeq{def:scale}.
    Using $\HQF^{(S=1)}(\beta_1)\propto \hat{H}_{\AKLT}^{(S=1)} $,
    the $S=1$ data (solid line) was obtained from a table in a previous study\cite{PRB.88.245118}
    and the $S=1$ data points without a solid line were calculated by DMRG additionally.
  }
\label{figGp}
\end{figure}

\Fig{figGp} shows the Haldane gap $\Delta E_+=E_+$ obtained using DMRG for a large system-size $N$.
In \fig{figGp}, one can find the system-size independence of each $S$ for $N>10$ and the small spin-$S$ dependence of ${E_+\over E_{+,N=4}}$ including $S=1$.
For $S=1$, the previous study\cite{PRB.88.245118} reported that the Haldane gap was $\Delta E_+ \simeq 0.350$,
which corresponds to $E_+/E_{+,N=4}=1.05$ in \fig{figGp} because the $S=1$ AKLT Hamiltonian \refeq{def:H_AKLT:P} has the energy unit $E_{+,N=4}=1/3$.
Another previous study\cite{PRB.106.134304} reported $2 \Delta E_+ \simeq 0.71$, i.e., $E_+/E_{+,N=4}= 1.065$, which seems to be an over estimation relative to \fig{figGp}.

Based on the above results, we conclude the existence of the Haldane gap at $q=\pi$,
\begin{eqnarray}
  \Delta E_+=\Delta E_{+,q=\pi} > 0
,
\end{eqnarray}
as an answer to the third question in \fig{figV}~(b).

\subsection{Goldstone-type Gapless Excitation\label{sec:Gm}}
The remaining question in \fig{figV}~(b) is about $\Delta E_-$.
In this subsection,  we clarify the nature of a one-magnon branch $\Delta E_{-,q}$. 
Before we discuss the one-magnon branch in the ferromagnetic AKLT model, we summarize the nature of the magnon in conventional ferromagnets.
In general, for fully-saturated ferromagnetic ground states with the total spin $S_\tot=N S$, $(2 N S+1)$-fold degeneracy reflects the spatial rotational symmetry $O(3)$ on spontaneous magnetization; one can spatially rotate magnetic moments with infinitesimally small energy for infinite system size.
Then, there is gapless excitation with a small wave number $q$, which corresponds to a spin twist with long distance, i.e., the Goldstone mode.
In fact, for ferromagnetic Heisenberg chains, the gapless one-magnon dispersion $\Delta E_{-,q}\propto 1-\cos{q}$ is obtained exactly\cite{ISBN.0521551439}.
For small $q$, the low energy excitation is approximated as  $\Delta E_{-,q} \propto q^2$, as depicted in \fig{figV}~(a).
Except for $q=0$, the low energy excitation $\Delta E_{-,q} $  has a total spin $S_\tot = N S -1$, which is decreased by one spin from a fully-saturated total spin $S_\tot=N S$.

\begin{figure}
  \begin{center}
  \includegraphics[width=0.52\textwidth]{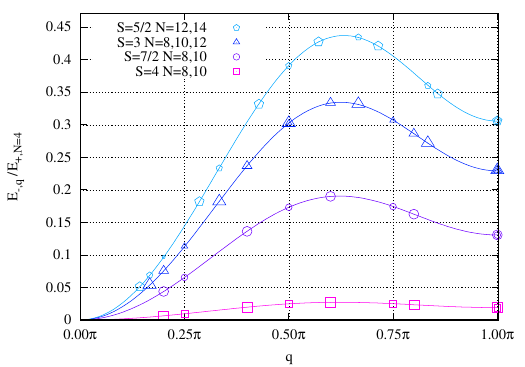}\\
  \end{center}
  \caption{(Color online)Low energy excitation $\Delta E_{-,q}=E_{-,q}$ as a function $q$ with $\Delta S=-1$ ($S_{\tot}=N(S-1)-1$), obtained using the Lanczos method.
    For each $S$, larger point indicates a larger system size $N$.
    The solid lines are fitted curves using the dispersion function in \refeq{def:DeltaE-}.
    $E_{+,N=4}$ is the unit of energy in \refeq{def:scale}.
  }
\label{fig3L}
\label{figGmL}
\end{figure}

\begin{figure}
  \includegraphics[width=0.52\textwidth]{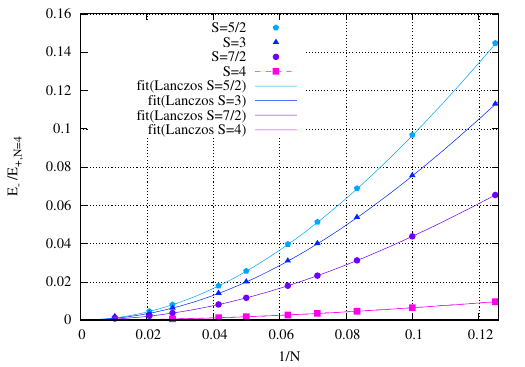}
  \caption{(Color online)System size $N$ dependence of the projected energy $E_{S_\tot=N(S-1)-1}$ in \refeq{ev:E_Stot}, obtained using DMRG.
    Solid lines are $E_{-,q=2\pi/N}$ of the fitted dispersion curves that were determined in \fig{figGmL} without using the DMRG results.
    $E_{+,N=4}$ is the unit of energy in \refeq{def:scale}.
  }
\label{figGm}
\end{figure}

In the ferromagnetic AKLT model, the lowest energy excited state in the sector $S_\tot^z = N(S-1)-1$ has a total spin $S_\tot= N(S-1)-1$ ($\Delta S=-1$), except for very small system-sizes.
Low energy excitation $\Delta E_{-,q}=E_{-,q}$ as a function of $q$ with $\Delta S=-1$ ($S_{\tot}=N(S-1)-1$) are obtained by the Lanczos method as shown in \fig{fig3L}.
The dispersion $E_{-,q}$ is approximated very well by fourth-order cosine bands with the fitting parameters $v_n$ defined as 
\begin{eqnarray}
  \Delta E_{-,q} = E_{-,q} = \sum_{n=1}^4 v_n (1-\cos{n q})
  \label{def:DeltaE-}
,
\end{eqnarray}
where the system-size independence for a small system size $N\leq 14$ is shown in \fig{fig3L}.
Through DMRG with \refeq{ev:E_Stot}, the lowest energy in the sector $S_{\tot}=N(S-1)-1$ is obtained for a larger system size $N$.
Although wave number $q$ is not obtained in DMRG, the fitted dispersion curves that were determined in \fig{figGmL} show good agreement with the DMRG data in \fig{figGm}.
Then, it can be expected that the lowest energy in the sector $S_{\tot}=N(S-1)-1$ has $q=2\pi/N$, even for a large system size $N$.

Based on the small $q$ expansion of \refeq{def:DeltaE-}, we conclude that the answer to the second question in \fig{figV}~(b) is
\begin{eqnarray}
  \Delta E_- \propto q^2
.
\end{eqnarray}
It should be emphasized here that the gapless excitation $\Delta E_-$ has $\Delta S=-1$ and does not exist for $S=1$ because the $S=1$ AKLT ``antiferromagnetic'' state has a total spin $S_\tot=0$ and there is no possibility for $\Delta S=-1$.
In other words, the existence of $\Delta E_-$ is a specific feature of ferromagnets---not only for quantum ferromagnets but also for classical ferromagnets---as shown in \fig{figV}.
In this sense, we call the dispersion of  $\Delta E_-$  a Goldstone-type gapless mode.

\subsection{Summary of Calculations\label{sec:Ea}}
As a short summary of this section, we have answered all three questions in \fig{figV}~(b): 1) unique total spin $S_\tot=N(S-1)$ of the ground states under the finite size gap, 2) the first excitation energy with $q={2\pi/N}$ comes from the Goldstone-type gapless mode $\Delta E_{-}\propto q^2$, and 3) a generalized Haldane gap $\Delta E_{+}>0$.
In general, however, many low energy states exist.
For example, there is the multi-magnon dispersion of the Goldstone-type gapless mode, which is considered to be a generalization of $m$-magnon modes $\Delta E_{-m}\propto (1-\cos{q})/m$ for spin-1/2 ferromagnetic Heisenberg chains\cite{ISBN.0521551439}.

\begin{figure}
  \begin{center}
  \includegraphics[width=0.52\textwidth]{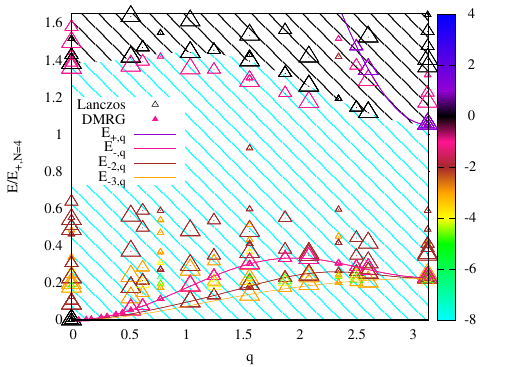}
  \end{center}
  \caption{(Color)Eigenenergy $E$ as a function of wave number $q$ for spin $S=3$, depicted with total-spin shift $\Delta S$ from -8 to 4 in the color scale.
    Larger point of the Lanczos data indicates larger system size $N$ in $N=8, 10, $ and 12.
    $E_{+,q}$ is the Haldane gapped dispersion and $E_{-m,q}$ is the Goldstone-type $m$-magnon dispersion.
    $E_{-,q}=E_{-1,q}$ is the same data in \fig{figGmL} and \fig{figGm}.
    In the light-blue shaded region, there should be many excitations with
    negative $\Delta S$,
    which are not captured numerically.
    $E_{+,N=4}$ is the unit of energy in \refeq{def:scale}.
  }
\label{figEa}
\end{figure}
\Fig{figEa} shows all the low energy states obtained using the Lanczos method and DMRG for $S=3$ .
The fitted lines of the Haldane gapped dispersion $E_{+,q}$ and the Goldstone-type $m$-magnon gapless dispersion $E_{-m,q}$ with $\Delta S=-m$ are also depicted in \fig{figEa}, where these were obtained by fitting \refeq{def:DeltaE-} with the lowest energies at a fixed $q$ in the fixed sector $S_\tot^z=N(S-1)-m$, as calculated using the Lanczos method for $m\leq 3$.
The dispersion $E_{-m,q}$ for $m\geq 4$ must exist but its $q$-dependence is not captured by the Lanczos method within $N\leq 12$.
Note that, for a fixed system size $N$, since the smallest wave number of the $m$-magnon mode is $q=2\pi m /N$, there exists the relationship $E_{-,q=2\pi/N} <E_{-2,q=4\pi/N}<E_{-3,q=6\pi/N}$; thus, the first excitation energy becomes $E_{-,q=2\pi/N}$, except for a very small system size $N$.

The existence of $E_{-m,q}$ for a large $m$ means there are more low energy excitations for a small $q$.
In fact, the Goldstone-type $m$-magnon gapless dispersion shows $E_{-,q} >E_{-2,q}>E_{-3,q}$ at a fixed $q>0$.
Then, in \fig{figEa}, we shade the low energy region with negative $\Delta S$ in light blue.
Similarly, in the light-blue shaded region for $0.6 \lesssim E/E_{+,N=4} \lesssim 1.4$, there should be many excitation data points with a small total spin, and there are not captured by the Lanczos method.
In other words, for $S_\tot \geq N(S-1)+1$ ($\Delta S \geq 1$), there is no uncaptured state below the Haldane gap $E/E_{+,N=4}\simeq 1.053$.
This means that there is a gapped structure for $S_\tot \geq N(S-1)+1$ in the light-blue shaded region, which plays an important role under a magnetic field, as discussed in the next section.

\section{Energy Gap Controlled by Magnetic Field\label{sec:Eb}}
In this section, 
we introduce a magnetic field term into the Hamiltonian as
\begin{eqnarray}
  \HQF^{(S)}(\beta_S) - h \hat{S}_\tot^z
.
\end{eqnarray}
The finite magnetic field $h$ splits the $(2S_\tot+1)$-fold degenerated eigenenergy by the addition of Zeeman energy $- h S_\tot, - h S_\tot+h, - h S_\tot+2h,\ldots, + h S_\tot$; this is the Zeeman splitting.
If a full energy diagram is classified by total spin $S_\tot$ under a zero magnetic field $h=0$, one can depict the energy diagram at finite $h$ without any calculation.

\begin{figure}
  \includegraphics[width=0.52\textwidth]{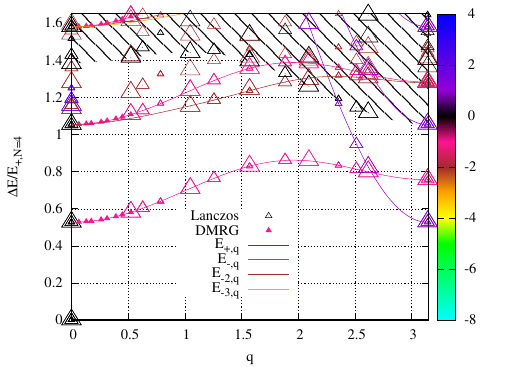}
  \caption{(Color online)Effect of Zeeman splitting on \fig{figEa} when the magnetic field $h$ is equal to half of the Haldane gap $E_{+,\pi}$, where the ground state is $\ketVBS$ in \refeq{VBS} and the ground state energy is $E_0= -h N(S-1)$.
  }
\label{figEb}
\end{figure}
\Fig{figEb} shows the energy diagram for spin $S=3$ when the magnetic field $h$ is equal to one half of the Haldane gap $E_{+,\pi}$.
Compared with \fig{figEa} covered with the light-blue region, the structure of low energy excitation is clarified with a finite gap.
In the following, we discuss this clarification induced by the magnetic field.

\begin{figure}
  \includegraphics[width=0.45\textwidth]{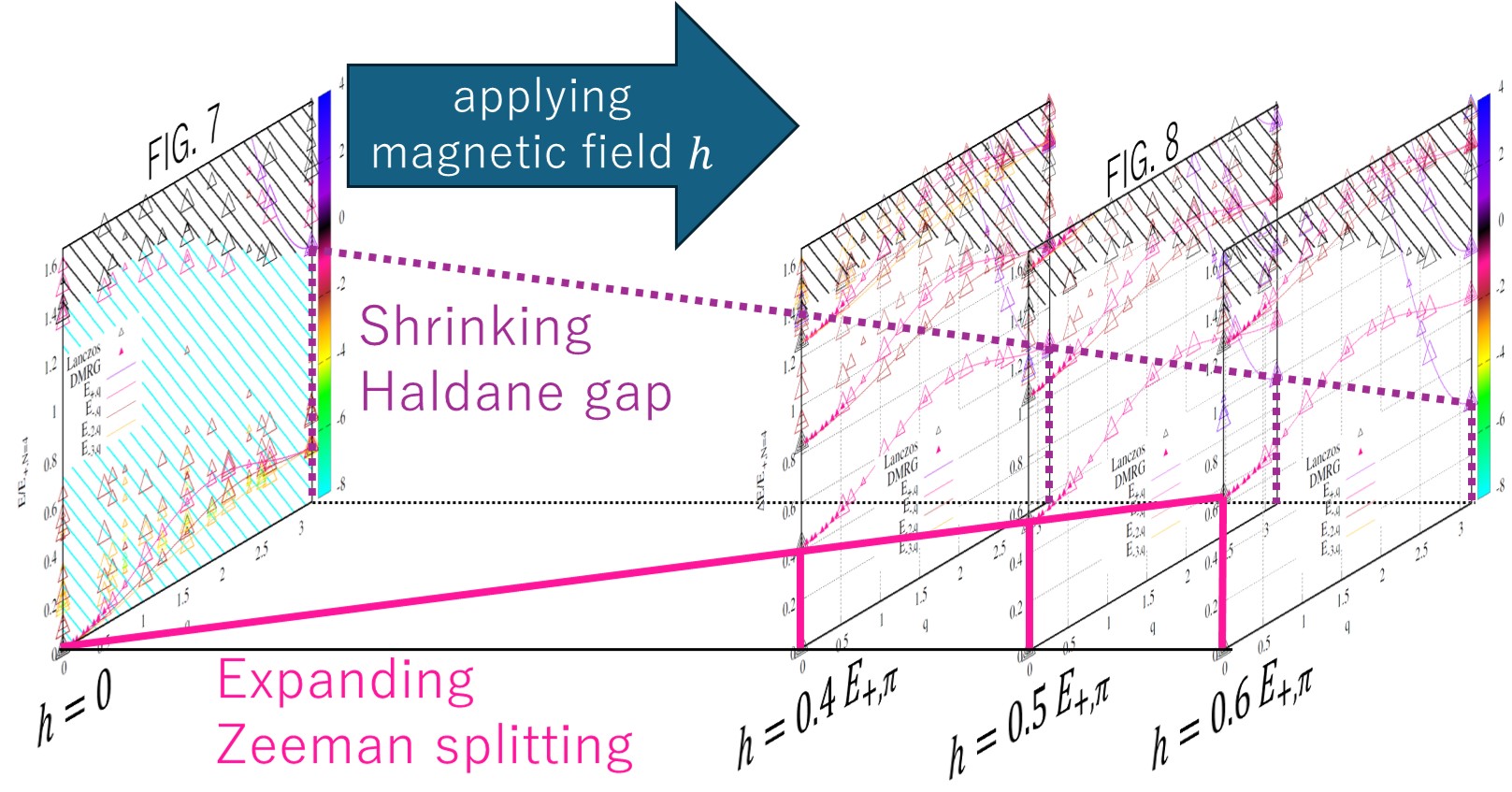}
  \caption{(Color online)Schematic figure of the energy diagram as a function of the magnetic field $h$.
  }
\label{figEbx}
\end{figure}
For Goldstone-type $m$-magnon gapless excitations $E_{-m,q}$ which were connected to $E=0$ at $q=0$, as shown in \fig{figEa},
the lowest excitation energy with Zeeman energy becomes
\begin{math}
  \Delta E_{-m,q} = E_{-m,q} - h [N(S-1)-m] + hN(S-1)= E_{-m,q} + m h
  \nonumber
.
\end{math}
This result indicates the emergence of the gap $m h$ for the gapless mode $E_{-m,q}$.
In fact, the lowest $E_{-,q}$ dispersion in \fig{figEb} has $\Delta S_\tot^z=-1$ and is connected to $\Delta E=h={E_{+,\pi}\over 2}$ at $q=0$.
In addition, both the second lowest $E_{-,q}$ dispersion and the lowest $E_{-2,q}$ dispersion have $\Delta S_\tot^z=-2$ and are connected to $\Delta E=2 h=E_{+,\pi}$ at $q=0$.
In this way, a state with a small $S_\tot < N(S-1)$ gains a large gap due to the Zeeman splitting(\fig{figEbx}).
Using the Lanczos method, we calculated a small enough $S_\tot^z$ sector to depict \fig{figEb}.

For the Haldane-gap branch with $E_{+,q}$ and $\Delta S=1$,
the lowest excitation energy with Zeeman energy is
\begin{math}
  \Delta E_{+,q} = E_{+,q} - h [N(S-1)+1] + h N(S-1)= E_{+,q} - h
  \nonumber
,
\end{math}
which indicates the shrinking of the Haldane gap $E_+=E_{+,q=\pi}$.
Then, it is expected that a transition from the original ground state $\ketVBS$ with $\Delta S=0$ and $q=0$ to the Haldane-gap state with $\Delta S=1$ and $q=\pi$
occurs under the finite magnetic field
\begin{math}
  h \geq  E_{+,\pi}
\end{math}
.
To confirm this transition, the ground state energy under the magnetic field $h$ is calculated by DMRG without considering the $S_\tot^z$ preservation.
As a result, the transition point is $h_c=E_{+,\pi}$ for a system size up to $N=24$, as expected.
In other words, there is a stable magnetic plateau of $S_\tot=N(S-1)$ for $|h|<h_c$.
The magnetic transition at $h_c=E_{+,\pi}$ from $\Delta S=0$ to $\Delta S=1$ means that a gapped structure also exists for high energy sectors with $\Delta S>1$.
If a state with $\Delta S$ has energy $E$ at $h=0$, its excitation energy at $h=h_c=E_{+,\pi}$ is
\begin{math}
  \Delta E = E - h_c [N(S-1)+\Delta S] - h_c N(S-1)
  = E - E_{+,\pi} \Delta S
.
\end{math}
Then, the above magnetic transition requires $E > E_{+,\pi} \Delta S$ for $\Delta S>1$, which is the gapped structure that is required to stabilize the ground state of $\Delta S=0$ in its magnetic plateau.
The situation is the same for the $S=1$ case, which has been established both theoretically and experimentally\cite{JMMM.140.1595}.
The major difference from the $S=1$ case is the controllability of lowest-excitation type; the lowest excitation comes from the one-magnon branch, $E_{-,q}+h$, for $|h|<E_{+,\pi}/2$ and the Haldane-gap branch, $E_{+,q}-h$, for $|h|>E_{+,\pi}/2$.
Such a controllable coexistence is common with that of the magnetization plateau state in spin systems.
Especially, ferrimagnetism exhibits similar combination of gapless ferromagnetic and gapped antiferromagnetic branches studied by Yamamoto, et al.\cite{JPSJ.67.3711, JPCM.10.11033, PRB.59.1024}, and its fractional magnetization is called as the Haldane plateau by Sakai and Okamoto\cite{PRB.65.214403}.
It should be noted that ferrimagnetism is explained by the Lieb-Mattis theorem\cite{JMP.3.749} but quantum ferromagnet is not.

To summarize this section, the low energy spectrum revealed by the magnetic field indicates the stable magnetic plateau
\begin{eqnarray}
  m={S_\tot \over NS}={S-1 \over S},  \;\; |h|<h_c=E_{+,\pi}
  ,
\end{eqnarray}
where we have the gapped and unique ground state $\ketVBS$.

\section{Application to MBQC\label{sec:MBQC}}
As the gapped and unique ground state $\ketVBS$ was established in \sec{sec:Eb}, a generalization of MBQC for the spin-$1$ AKLT model\cite{PRL.101.010502} is straightforward.
Four-fold degenerated ground states under the open boundary condition (OBC), i.e., $\HQF^{(S)}|_{J^{(s)}_N=0}$ in \refeq{eq:FV}, are written in the MPS as
\begin{eqnarray}
  \ket{\vec{L},\vec{R}}_{[1:N]}=\trans{(L_0,L_1)}\prod_{i=1}^N A_i \vv{R_0}{R_1}
  .
\end{eqnarray}
The left and right edge states can be determined as two-dimensional complex vectors $\vec{L}=\vv{L_0}{L_1}$ and $\vec{R}=\vv{R_0}{R_1}$.
One key idea in MBQC is that the unitary transformation of a two-dimensional complex ``q-bit'' vector can be realized by projection measurement onto the corresponding edge state.
In fact, using an appropriate basis $\ket{p_m}$, $(m=1,\ldots 2S+1)$, the projection measurement of the $N$-th site on the right-edge can generate a new state $\ket{\vec{L},U_m \vec{R}}_{[1:N-1]} \ket{p_m}_N$ with a determined state $\ket{p_m}_N$ at the $N$-th site, where $U_m$ is a two-dimensional matrix and $\vec{R}$ is the q-bit vector.

As a generalization of $S=1$ MBQC, we define the (unnormalized) orthogonal basis
\begin{eqnarray}
  \ket{p_1}&=&\ket{S}-\sqrt{S(2S-1)}\ket{S-2}+C \sqrt{S-1}\ket{S-3}
  ,\nonumber\\
  \ket{p_2}&=&-\i \ket{S}-\i \sqrt{S(2S-1)}\ket{S-2}
             \nonumber \\
  &&-\i {(2S+1)\sqrt{S-1}\over C}\ket{S-3}
  ,\nonumber\\
  \ket{p_3}&=&-\sqrt{2S}\ket{S-1}
             ,\label{eq:mbqc}
\end{eqnarray}
with a free parameter $C$, which is required for $\braket{p_1}{p_2}=0$.
Here, for $S=1$, $\ket{S-3}$ becomes non-physical but does not appear due to the zero coefficient, $\sqrt{S-1}=0$.
For $m=1,2,$ and 3, the corresponding unitary matrices $U_1, U_2, U_3$ become the Pauli matrices as
\begin{eqnarray}
  U_1=
  \left(
  \begin{array}{cc}
    0 & 1 \\
    1 & 0 \\
  \end{array}
  \right)
  ,
  \nonumber\\
  U_2=
  \left(
  \begin{array}{cc}
    0 & -i \\
    i & 0 \\
  \end{array}
  \right)
  ,
  \nonumber\\
  U_3=
  \left(
  \begin{array}{cc}
    1 & 0 \\
    0 & -1 \\
  \end{array}
  \right)
,
\end{eqnarray}
whose proof is simply a calculation on
\begin{eqnarray}
 U_m=\bra[N]{p_m}A_N
  =
   \left(
  \begin{array}{cc}
    -{\braket[N]{p_m}{S-1}\over \sqrt{2S}}&\braket[N]{p_m}{S} \\
    - {\braket[N]{p_m}{S-2} \over \sqrt{S(2S-1)}} &{\braket[N]{p_m}{S-1}\over \sqrt{2S}} \\
  \end{array}
  \right)
  \nonumber
.
\end{eqnarray}
This is a direct generalization of spin-$1$ MBQC\cite{PRL.101.010502}.
For $S>1$, other generalizations with using $\ket{m}$ $(m<S-3)$ can be possible and might be suitable for measurement along rotated spin axis.

Compared with the spin-1 MBQC, a finite magnetic field, which is not required for the spin-$1$ MBQC, is required to realize the gapped and unique ground state.
The magnetic field also affects edge states, which can be a drawback.
However, there can be new properties which do not exist in the spin-$1$ antiferromagnetic MBQC, such as the magnetic field control of spontaneously-magnetized ground states, inter-edge interaction through the Goldstone-type one magnon modes, edge states at domain-wall boundaries.

\section{Summary\label{sec:summary}}
In summary, we have generalized the spin-1 AKLT model to spin-$S$ ferromagnetic AKLT models following the previous studies\cite{JPCM.4.7469, PRB.82.104404}, and we have presented an application of the quantum ferromagnets to MBQC.
Except for $S=3/2$ and $S=2$, the ground states of the models are the quantum ferromagnets and have unique reduced ferromagnetic moment $S_\tot=N(S-1)$, i.e., the fractional magnetization $m=(S-1)/S$, even under a zero magnetic field.
In short, this quantum ferromagnet is a ``magnetic chimera'' of a classical ferromagnet and quantum antiferromagnet, as shown in \fig{figV}.
Reflecting the coexistence of the two types of magnets, the low energy excitations are composed of both gapless $\Delta E_- \propto q^2$ and gapped $\Delta E_+>0$; the former is a one-magnon mode of the Goldstone-type multi-magnon gapless modes and the latter is a generalization of the Haldane gap. The gapless mode $\Delta E_-$ are characteristic to ferromagnets, whereas the gapped mode $\Delta E_+$ is an antiferromagnetic character of Haldane spin chains.
The magnetic chimera having the ability to break the spin rotational symmetry spontaneously and the potential for application to MBQC, hidden behind continuous magnetic excitations under a zero magnetic field, appears clearly when a magnetic field is applied.
Just as the original AKLT model has been a rigorous starting-point to explore a wide class of SPT phases theoretically and Haldane materials experimentally, the ferromagnetic AKLT models will play the same role in theoretical studies and experimental observations.

Contrary to spin-$S$ antiferromagnetic Heisenberg chain models, the Haldane gap $\Delta E_+>0$ in the spin-$S$ ferromagnetic AKLT models exists in both integer and half-integer spin-$S$ models.
On the other hands, the previous study on the quantum ferromagnets with magnetization $m=(S-1/2)/S$ in spin-$S$ BLBQ models\cite{PRB.108.140404, PRB.110.014433} shows the des Cloizeaux-Pearson mode ($\Delta E_- \propto |q|$) known in $S=1/2$ Heisenberg chain\cite{PRB.108.140404}, which does not depend on $S$, either.
In this way, we expect that the spin parity effect for the quantum ferromagnets does not depend on spin $S$ but depends on the macroscopic shrinking of $S_\tot$ as a variable $s=S-S_\tot/N$ as shown \fig{figV}~(b).
After we quantify the shrinking of $S_\tot$  in 
\begin{eqnarray}
 S_\tot=N (S-s) 
, 
\end{eqnarray}
{\it the spin-$s$ liquefaction of ferromagnetic moment due to quantum spin fluctuation} gives rise to des Cloizeaux-Pearson mode ($\Delta E_- \propto |q|$) for $s=1/2$\cite{PRB.108.140404} and the Haldane gap ($\Delta E_- >0$) for $s=1$, as shown in \fig{figV}.
This spin-parity effect that depends on the amount of spin liquefaction $s$ is completely different from the spin-$S$ dependent spin-parity effect on ferromagnets\cite{PRB.107.024403}.
In other words, the spin parity effect for the quantum ferromagnets, i.e., an extension of the {\it Haldane's conjecture}, should be discussed based on the shrinking of ferromagnets $s$ rather than the spin-$S$ in the models.
Because spin-$(s\geq 3/2)$ liquefaction has not been established yet, the generalized spin-parity effect is a conjecture at this stage.
Note that this conjecture for SU(2) symmetric ferromagnets is completely different from the SU($n$) generalization of the Haldane's conjecture\cite{NPB.952.114932}.

The ground state of quantum ferromagnets after the spontaneous symmetry breaking has quantum entanglement coming from the quantum entanglement in spin-$s$ antiferromagnets because the classical ferromagnetic background cannot contribute to quantum entanglement.
The above theoretical results not only abolish the prejudice that ferromagnetism is classical but also will open another frontier of ``quantum ferromagnetism'' in this new era of quantum computer science.

\begin{acknowledgments}
This work was supported by Japan Society for the Promotion of Science (JSPS) KAKENHI Grants No. JP22H01171.
The computation was partly carried out using the computer resources offered by the Research Institute for Information Technology, Kyushu University.
\end{acknowledgments}
 \bibliography{draft2.bib} 

\begin{thebibliography}{62}
\bibitem [{not({\natexlab{a}})}]{note1}%
  \BibitemOpen
  \href@noop {} {} ({\natexlab{a}}),\ \bibinfo {note} {detail of the
  ferromagnetic property, $\Delta E_- \propto q^2$, about spin-1/2 liquefaction
  will be published elsewhere to answer an checked question in the top panel of
  \fig{figV}~(b).}\BibitemShut {Stop}%
\bibitem [{not({\natexlab{b}})}]{note2}%
  \BibitemOpen
  \href@noop {} {} ({\natexlab{b}}),\ \bibinfo {note} {the symmetrization
  mapping operator is defined as $\symop=\prod_{i=1}^N \symop_i$ with
  $\symop_i=\sum_{m=-S}^S \ket[i\;\;i]{m}\bra{S,m}$, where $\ket[i]{m}$ is
  spin-$S$ local basis and $\ket[i]{S,m}$ is local basis of decomposed spins
  defined as $\ket[i]{S,S}=\ket{+1/2}_{i,R}\ket{+1/2}_{i,L}\ket{S-1}_{i,F}$ and
  $\ket[i]{S,m-1}={({\hat{s}}_{i,L}^-+{\hat{s}}_{i,R}^-+{\hat{s}}_{i,F}^-)\ket[i]{S,m}\over
  \sqrt{(S+m)(S-m+1)}}$ for $m\leq S$. This is a direct generalization of the
  $S=1$ case\citep{JMP.64.031901}. Using the definition, one can deduce
  \refeq{eq:symop}.}\BibitemShut {Stop}%
\end{thebibliography}
\end{document}